\documentclass[reprint,amsmath,amssymb,aip, apl, superscriptaddress]{revtex4-1}

\usepackage{graphicx}
\usepackage{bm}
\usepackage{gensymb}
\usepackage{hyperref}

\begin{document}

\preprint{APS/123-QED}

\title{Oscillatory Exchange Bias Controlled by RKKY in Magnetic Multilayers}

\author{D.~M.~Polishchuk}
\affiliation{Nanostructure Physics, Royal Institute of Technology, 10691 Stockholm, Sweden}
\affiliation{Institute of Magnetism of NAS and MES of Ukraine, 03142 Kyiv, Ukraine}
\author{M.~Persson}
\affiliation{Nanostructure Physics, Royal Institute of Technology, 10691 Stockholm, Sweden}
\author{M.~M.~Kulyk}
\affiliation{Nanostructure Physics, Royal Institute of Technology, 10691 Stockholm, Sweden}
\author{G.~Baglioni}
\affiliation{Nanostructure Physics, Royal Institute of Technology, 10691 Stockholm, Sweden}
\author{B.~A.~Ivanov}
\affiliation{Institute of Magnetism of NAS and MES of Ukraine, 03142 Kyiv, Ukraine}
\affiliation{Institute for Molecules and Materials, Radboud University, Nijmegen, the Netherlands}
\author{V.~Korenivski}
\affiliation{Nanostructure Physics, Royal Institute of Technology, 10691 Stockholm, Sweden}

\begin{abstract}
Ferromagnetic/antiferromagnetic bilayers are interfaced with normal metal/ferromagnetic bilayers to form F*/AF/N/F valves. The N-spacer thickness is chosen such that it mediates strong indirect exchange (RKKY) between the outer ferromagnetic layers, which varies in strength/direction depending on the N thickness and in direction on switching F. The system exhibits a strong modulation of the F*/AF exchange bias, oscillating in strength synchronously with the oscillation in the interlayer RKKY exchange across the normal metal spacer. The effect is explained as due to a superposition taking place within the antiferromagnetic layer of the direct-exchange proximity effect from the F*/AF interface and the indirect RKKY exchange from F penetrating AF via N. The modulation, expressed via the strength of the F*/AF bias field, reaches 400\% at the first RKKY peak.
\end{abstract}

\maketitle

Exchange biasing ferromagnetic films by interfacing them with antiferromagnetic layers is a widely used mechanism for producing unidirectional anisotropy in magnetic multilayered materials used in various technological applications~\cite{Nogues1999,Nogues2005,Hellman2017}. Exchange bias in a F*/AF bilayer is essentially fixed in fabrication and is non-trivial to control in case an application requires variable anisotropy. One notable exception is the thermally-assisted magnetic random access memory type structures, where the strength of the exchange bias is controlled by varying the device temperature across the Néel point of the AF in a F*/AF bilayer~\cite{Prejbeanu2007,Prejbeanu2013}. The useful effect is achieved by modulating the strength of the antiferromagnetic order in AF, which in turn modulates the pinning strength of the F-layer, by using thermal agitation of the material within the device volume.

\begin{figure}
\includegraphics{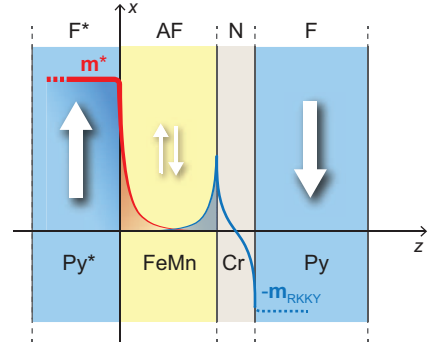}
\caption{\label{fig:idea} Physical layout of multilayer system F*/AF/N/F (Py/FeMn/Cr/Py) under study: induced direct (by F*) and indirect (by F) exchange ($\mathbf{m}^*$, $\mathbf{m}_\text{RKKY}$) penetrate and affect magnetic state of AF, where indirect exchange can be modulated during and post-fabrication.}
\end{figure}

An interesting question is weather the AF-order can be affected in a more focused way, electromagnetically, at any temperature/operating point. The candidate effects to consider would be the finite-size~\cite{Hernando1995,Navarro1996,Kravets2012} and exchange-proximity~\cite{Offi2002,Lenz2007,Golosovsky2009} effects, which are often atomically local and strong enough to compete with the intrinsic exchange in the AF layer. Our approach is to use indirect exchange produced by a N/F bilayer~\cite{Gruenberg1987,Baibich1988,Parkin1990} in the strong-RKKY limit (ultrathin N) that can penetrate via the AF/N interface and superpose within the AF layer with the direct-proximity exchange from the biased F*/AF interface in a F*/AF/N/F multilayered structure (depicted in Fig.~\ref{fig:idea}). The  result is a modulation of the magnetic ordering in AF. We show experimentally and explain theoretically that such RKKY-control, which can be varied in strength and direction during fabrication as well as dynamically post-fabrication, results in a pronounced oscillation of the F*/AF exchange bias versus the N-spacer thickness, with the effective modulation reaching a factor of 4 at the first RKKY peak.

\emph{Samples and methods.} The multilayered system F*/AF/N/F used in this work is illustrated in Fig.~\ref{fig:idea} and has the following material parameters:  Py*(10 nm)/FeMn(6 nm)/Cr($t_{\text{Cr}}$)/Py(20 nm), $t_{\text{Cr}} = 0, 0.5, ..., 5$ nm. Permalloy (Py) is used owing to its easy switching properties as well as it being a good buffer layer for obtaining the optimal properties of the antiferromagnetic FeMn. The difference in thickness of the two Py layers makes it easy to distinguish them using magnetometry. The thickness of the FeMn layer is chosen to be small enough to exhibit strong finite size effects~\cite{Lenz2007,Polishchuk2021a}, such that the magnetic state of its interfaces would significantly affect its AF ordering; 6~nm was found to be optimal in our case. The normal metal spacer (N) thickness was varied through the entire RKKY range, to investigate the effect of the oscillating indirect exchange mediated by the conduction electrons in N (RKKY) on the AF-ordering strength in the FeMn layer. Switching of the right Py layer (F) in this configuration reverses the sign of the RKKY, which is expected to superpose constructively or destructively with the direct-exchange propagating into the AF layer from the left ferromagnetic interface (magnetic proximity effect). Samples with negligible RKKY coupling (Cr spacer thickness $\gtrsim$ 3~nm) were fabricated and served to calibrate the RKKY-effect in focus. 

All samples were deposited on undoped Si (100) substrates at room temperature using a UHV dc magnetron sputtering system (AJA Inc.). Before deposition, the substrates were Ar ion-milled in 5~mTorr Ar atmosphere by applying a 45-W RF bias for 10 min. The thicknesses of the individual layers were controlled by setting the deposition time using the respective rate calibrations.

Vibrating-sample magnetometry (VSM) was used to perform the in-plane magnetization measurements at room temperature (Lakeshore Inc.). The ferromagnetic contribution was refined by subtracting the background, which was measured for the bare Si substrates of the same volume as those carrying the multilayers. Cavity-FMR measurements were carried out at a constant operating frequency of 9.36~GHz using an X-band ELEXSYS E500 spectrometer (Bruker Inc.). Broadband FMR measurements were performed using a custom-made coplanar waveguide setup, with a variable-frequency GHz generator and a microwave diode (both Agilent Inc.), locked-in to the modulated field of the biasing electromagnet. 

\begin{figure}
\includegraphics{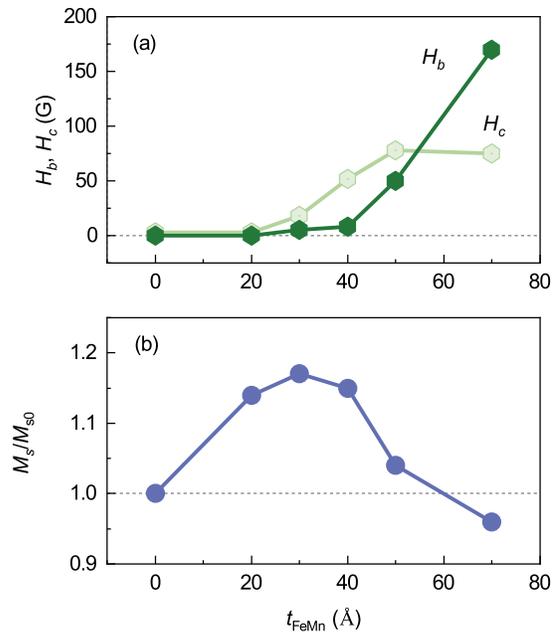}
\caption{\label{fig:Ref} (a) Exchange bias field, $H_b$, and coercivity field, $H_c$, measured for reference Py(6~nm)/FeMn($t_{\text{FeMn}}$) bilayers with varying $t_{\text{FeMn}}$. (b) Changes in saturation magnetization, $M_s$, of reference bilayers with respect to single 6-nm Py film, $M_{s0}$.}
\end{figure}

\emph{Competing interactions.} In the antiferromagnetic FeMn layer, in the vicinity of the F*/AF interface, the intrinsic antiferromagnetic exchange is in competition with the ferromagnetic proximity-induced exchange due to the direct interlayer exchange interaction at the interface. For very thin AF layers, 4~nm and thinner, the AF-order is fully suppressed as expressed by the measured zero bias field of the F* layer in this regime; shown in Fig.~\ref{fig:Ref}(a) for a control experiment on Py*(6 nm)/FeMn($t_{\text{FeMn}}$) bilayers. For thicker AF layers, the AF-order is recovered and the bias field becomes finite. The corresponding expected proximity-induced magnetization profile for this intermediate regime of near-critical AF thickness (4-6~nm) is depicted in Fig.~\ref{fig:idea} in red. Fig.~\ref{fig:Ref}(b) shows the directly measured induced ferromagnetic polarization of the thin FeMn layers, which confirms the presence of proximity-induced magnetization in AF, with a pronounced nonlinear dependence on $t_{\text{FeMn}}$. The FeMn thickness of 6~nm, selected to be the focus in this paper, corresponds to the strongest competition between the intrinsic AF order and the proximity effect at room temperature. \footnote{Interestingly, the data in Fig.~\ref{fig:Ref}(b) indicate that FeMn thicker than 6 nm is strong enough to slightly depolarize the Py* surface -- its normalized magnetization is slightly below 1; although more samples and measurements are needed to ascertain this initial indication from a single data point.}

Much of the strength in the AF-order for 6 nm thick FeMn comes from the outer region (near its right surface), which is weakly affected by the F*-proximity effect. Our idea is to use additional exchange focused onto this region in order to modulate the AF-order. It would be most interesting if this additional exchange could be modulated in both strength and direction, such that it could combine with the F*-proximity exchange to diminish or, in fact, enhance the AF-order in FeMn. The most natural candidate is indirect exchange (also known as RKKY exchange) from a ferromagnetic layer via an ultra-thin normal metal spacer. Its extra advantage is that there is no direct exchange coupling AF-to-F via N and therefore the F layer can be easily switched to alter the direction of the RKKY. This would then alter the superposition of the proximity and the RKKY contributions within the AF layer and be expected to modulate the AF-order and, consequently, the exchange bias of F*.

\begin{figure}
\includegraphics{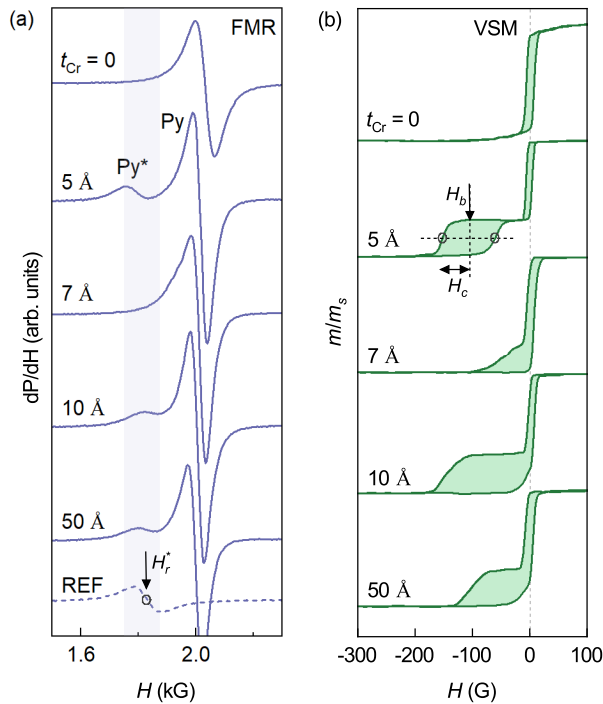}
\caption{\label{fig:FMR} FMR and VSM data for Py*/FeMn/Cr/Py multilayers. (a) In-plane FMR spectra obtained at 14~GHz with external field applied along pinning direction (0\degree) and (b) VSM hysteresis loops, for select structures with $t_{\text{Cr}} = 0, 5, 7, 10, 50$~\AA. Dashed line in panel (a) is FMR spectrum for reference bilayer sample of Py*(10~nm)/FeMn(6~nm).}
\end{figure}

\emph{Magnetometry and FMR.} The structure with $t_{\text{Cr}}=5$~nm has no interlayer coupling since RKKY vanishes at about 3~nm of the Cr spacer thickness. As a result, the Py*/FeMn bilayer and the Py layer (F) are independent magnetically and the corresponding FMR spectrum and VSM loop, Fig.~\ref{fig:FMR}(a) and (b), consist of two distinct contributions from the pinned Py* and free Py layers.

The resonance line for the reference bilayer Py*(10~nm)/FeMn(6~nm), dashed in Fig.~\ref{fig:FMR}(a), shows the position of the resonance field $H_r^*$ of the pinned Py* layer in the limit of no-to-weak RKKY-exchange, throughout the sample series. $H_r^*$ is lower than $H_r$ of the free Py layer owing to the additional anisotropy from the F*/AF exchange biasing.

\begin{figure}
\includegraphics{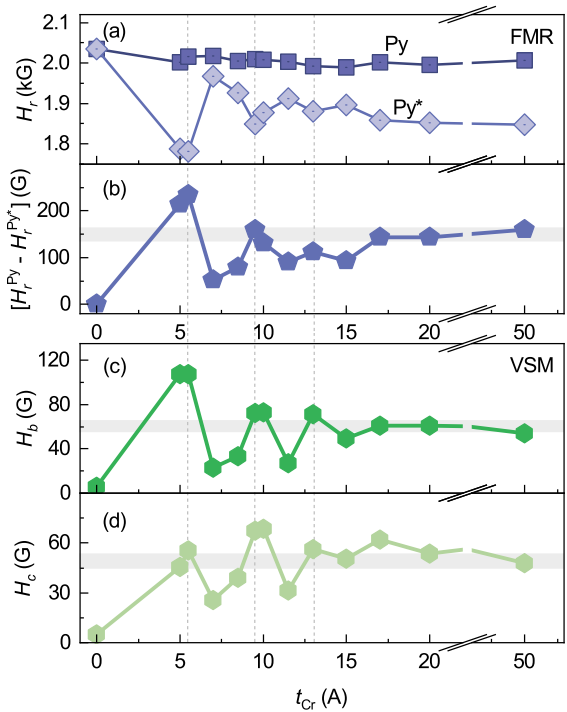}
\caption{\label{fig:VSM} (a) Resonance field of Py and Py* layers and (b) their difference at 14~GHz; (c) exchange-bias field $H_b$ and (d) coercivity field $H_c$ of Py* minor hysteresis loop; all versus Cr spacer thickness. See panel (b) of Fig.~\ref{fig:FMR} for definition of $H_b$ and $H_c$. Horizontal grey lines in (b)-(d) indicate respective base line values for vanishing RKKY (large $t_{\text{Cr}}$ limit).}
\end{figure}

On changing $t_{\text{Cr}}$, both the resonance line and the minor hysteresis loop of the Py* layer, shown for select samples in Figs.~\ref{fig:FMR}(a) and (b), exhibit changes in their positions. These are plotted in Figs.~\ref{fig:VSM}(a)-(d) versus the spacer thickness and show an oscillatory character with a period of about 0.6~nm in the Cr thickness.

The periodic changes in the extracted parameters, $H_r^*$, $H_b$, and $H_c$, reflect the modulation of the exchange bias in the Py*/FeMn bilayer by the oscillatory RKKY interlayer exchange coupling between the FeMn and the free Py layer through the Cr spacer, whose thickness is varied.  

\emph{Direct-proximity versus RKKY exchange.} The direct proximity effect from F* penetrating into AF is expected to decay in strength from the F*/AF interface. The characteristic decay length is a few nanometers, judging by the profile of the induced magnetization in AF, shown in Fig.~\ref{fig:Ref}(b). The proximity effect is significant over about 4 nm into the AF layer. 

In contrast to the proximity exchange due to the direct interaction of the lattice spins across the F*/AF interface, the RKKY exchange is carried by the spin-polarized conduction electrons, which should be expected to propagate (most likely maintaining a degree of spin polarization) at least as deep as the direct lattice-spin exchange. The two extrinsic-to-AF exchange contributions would then overlap and superpose within the AF layer. Naturally, this superposition would result in a weaker or stronger effective exchange field (of ferromagnetic origin, suppressing the AF order, as shown by Fig.~\ref{fig:Ref} and explained in the theory section below) depending on whether the superposition is constructive or destructive, i.e., depending on the sign (in addition to the strength) of the RKKY term, which in turn can change on varying the Cr-spacer thickness or switching the F layer's magnetic orientation. 

The observed oscillations in $H_b$ and $H_r$ correlate very well and both show that, importantly, the exchange bias can be decreased or increased over the base line corresponding to vanishing RKKY (found at about 60 Oe for $H_b$; increased by about two-fold to $H_b =$110 Oe at the $1^\text{st}$ RKKY peak). Only a decrease in $H_b$ would be expected if the direct-proximity and RKKY exchange profiles did not overlap (did not superpose) within the AF layer, since the action of the RKKY contribution would only suppress the AF order (at the AF/N interface, independently of proximity exchange). In the case of an overlap, the effective total ferromagnet-induced exchange should decrease for a destructive superposition (for the case of opposing signs of the direct-proximity and RKKY exchange terms), which indeed should enhance the overall AF order by the presence of RKKY and thereby lead to a stronger exchange bias compared to the no-RKKY base line. In other words, RKKY can act to reduce the suppression of the AF order by the proximity exchange, which leads to a stronger F*/AF exchange bias.

\begin{figure}
\includegraphics{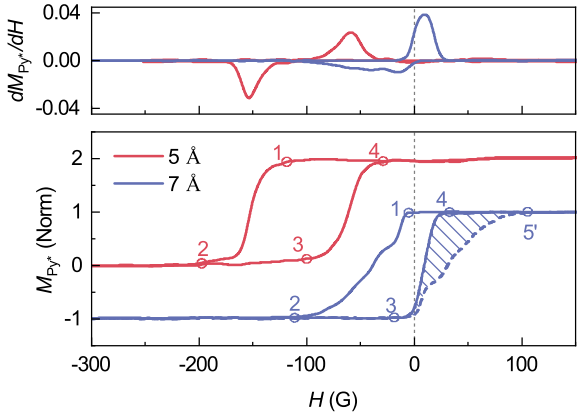}
\caption{\label{fig:ML} VSM minor hysteresis loops for two samples with $t_\text{Cr} =5$ and 7~Å. Top panel: field derivatives for respective loops. Data extracted from Fig.~\ref{fig:FMR}(b); see text for details. }
\end{figure}

\emph{Theoretical considerations.} Let us consider the system taking the effects of the F* and F layers on the AF layer to be independent of its magnetic state, i.e., of the magnitude of its antiferromagnetic vector \textbf{L}. It then follows that the two ferromagnets produce in AF an effective field $\textbf{H}_\text{eff}$, which is the sum of the exchange field \textbf{H*} connected with the proximity-induced magnetization due to F* and the RKKY exchange field $\textbf{H}_\text{RKKY}$ from F (via N). The orientations of the fields \textbf{H*} and $\textbf{H}_\text{RKKY}$ are determined by the respective magnetizations of the F* and F layers, which are collinear, parallel or antiparallel. For the AF layer, the order parameter is \textbf{L}, which is affected by $\textbf{H}_\text{eff}$. In the spirit of the phenomenological theory of antiferromagnetism~\cite{LL8}, we write the thermodynamic potential $\Phi (\textbf{L},\textbf{H}_\text{eff})$ as a function of the AF-vector \textbf{L}:
\begin{equation}
    \Phi = f(L^2) + D(\textbf{H} \cdot \textbf{L})^2 + D'H^2L^2 - \frac{\chi_p}{2}\textbf{H}^2.
\end{equation}
Here and below \textbf{H}=$\textbf{H}_\text{eff}$ for brevity of notations and $L=\left| \textbf{L} \right|$. The first term describes the energy of the AF in the absence of an external field, the next two terms with phenomenological constants $D>0$ and $D'>0$ determine the action of the field (note that the linear term $\textbf{H} \cdot \textbf{L}$ is forbidden by AF symmetry), and $\chi_p>0$ is the paramagnetic susceptibility of the AF; for details see~\cite{LL8}. 

Since $D>0$, one can consider $\textbf{H} \cdot \textbf{L} = 0$, then using $\delta \Phi/\delta H = -\textbf{M}$ we obtain $\textbf{M} = \chi \textbf{H}$, $\textbf{M} \perp \textbf{L}$, with the AF susceptibility to the effective field
\begin{equation}\label{eq2}
    \chi = \chi_p - 2D'L^2.
\end{equation}
This result means that the effective field induces an additional magnetic moment in AF. If the thickness of the AF layer is small, $L$ is suppressed and the induced moment is relatively large. The induced moment decreases as the AF thickness increases. 

The suppression of the AF order by the presence of a strong local exchange can be further illustrated using the Landau expansion~\cite{LL8}:
\begin{equation}
    f = -\frac{1}{2}a(T_N-T)L^2+\frac{b}{4}L^4,
\end{equation}
with $T_N$ being the Néel temperature of the AF, which yields for the magnitude of the AF order parameter
\begin{equation}\label{eq4}
    L^2=L_0^2\begin{cases}
    (1-H^2/H_c^2), & \text{$H<H_c$},\\
    0, & \text{$H \geq H_c$}.
  \end{cases}
\end{equation}
Here $L_0=\sqrt{a(T_N-T)/b}$ is the value of the AF order parameter in the absence of the field, and $H_c^2=a(T_N-T)/2D'$ has the meaning of a critical field at which the AF order is fully suppressed. Writing out the effective field explicitly, $H^2=(H^*)^2+(H_\text{RKKY})^2+2H^*H_\text{RKKY}$, one can see the key effect of sub-critical interface-induced exchange in our case. Namely, sub-critical exchange fields of any sign (any F*, F orientation) partially suppress the AF order, however, the magnitude of this suppression depends on the signs of these fields, i.e., on the relative orientation of the outer ferromagnetic layers F* and F. If the strengths of the two exchange fields are comparable and not too small, the induced change of the AF order parameter $L$ can be significant.

As shown above, the direct-proximity and RKKY exchange, significantly penetrating and overlapping in the AF layer, necessarily modulate the effective exchange field in AF, which in turn modulates the AF order parameter according to \eqref{eq2},\eqref{eq4}. The strength of the AF order then directly scales the magnitude of the exchange bias the layer can provide. As a result, the exchange bias field oscillates with the oscillating in magnitude and sign RKKY exchange -- a function of the thickness of the spacer and the direction of the free layer in the studied multilayer structure. 

\emph{Exchange bias vs F switching.} Figure~\ref{fig:ML} shows minor hysteresis loops for the two samples most representative of enhanced ($t_\text{Cr} =5$~Å) and suppressed ($t_\text{Cr} =7$~Å) exchange bias, showing qualitatively different behavior attributed to, respectively, static versus switching F layer: F layer's magnetization is static for the 5Å-sample, it switches when field crosses into the positive range for the 7Å-sample. Numerals mark four key states of mutual F*-F magnetization orientation: 1AP-2P-3P-4AP (red); 1AP-2P-3P-4P (blue). The top panel shows the field derivatives for the respective loops: nearly symmetric (5Å-sample, final state AP) versus clearly asymmetric in the case where the transition involves switching of the F layer (7Å-sample; final state P; 4-fold enhanced $dM/dH$ at $H>0$ on P to AP transition). Dashed line is the expected form of the right transition in the absence of F-switching (symmetric minor loop expected, see e.g. 5Å- and 10Å-sample data; point 5' is mirrored point 2-blue). The shaded area marks the difference with the measured loop of the pinned layer and is attributed to switching of the RKKY-controlling F layer.

\emph{In conclusion}, we show that the widely used in technology F*/AF exchange bias can be modulated by incorporating a N/F bilayer into the structure designed to affect the strength of the antiferromagnetic order in AF via the RKKY exchange interaction. The resulting F*/AF exchange bias shows an oscillatory behavior characteristic of RKKY as the N-spacer thickness is varied. The depth of the exchange-bias field modulation reaches a factor of 4 near the first RKKY peak. This effect should be interesting for applications where exchange bias needs to be controlled, such as magnetic memory and sensors.

\emph{Acknowledgments.} Support by the Swedish Research Council (VR\#2018-03526) and the Olle Engkvist Foundation (2020-207-0460) are gratefully acknowledged.

\bibliography{apssamp}

\end{document}